\documentclass{article}
\usepackage{emulateapj,pstricks,apjfonts}

\def\1wga{\mbox{1WGA~J1226.9$+$3332}}
\def\fxv{\mbox{$F_X/F_V$}}
\def\cgs{\mbox{erg cm$^{-2}$ s$^{-1}$}}

\begin{document}

\submitted{ApJ accepted}

\lefthead{1WGA~J1226.9$+$3332 : a high redshift cluster}
\righthead{Cagnoni et al.}

\title{1WGA~J1226.9$+$3332 : a high redshift cluster discovered by {\em Chandra}}

\author{I. Cagnoni\altaffilmark{1,2}, M. Elvis\altaffilmark{2}, D.-W. Kim\altaffilmark{2}, P. Mazzotta\altaffilmark{2,3}, J.-S. Huang\altaffilmark{2}
\and  A. Celotti\altaffilmark{1}}
\affil{$^{1}$ SISSA, Via Beirut 4 -34138, Trieste, Italy}
\affil{$^{2}$ Harvard-Smithsonian Center for Astrophysics, 60 Garden Street, 
Cambridge, MA 02138, USA}
\affil{$^{3}$ ESA fellow}
\footnote{email: Ilaria Cagnoni: ilale@sissa.it or icagnoni@cfa.harvard.edu\\
 Martin Elvis: melvis@cfa.harvard.edu\\ 
Dong-Woo Kim: dkim@cfa.harvard.edu\\ 
Pasquale Mazzotta: pmazzotta@cfa.harvard.edu\\ 
Jiasheng S. Huang: jhuang@cfa.harvard.edu\\ 
Annalisa Celotti: celotti@sissa.it}

\begin{abstract}

We report the detection of \1wga \/ as  an  arcminute scale 
extended X-ray source with the {\em Chandra} X-ray Observatory.
The {\em Chandra} observation and R and K band imaging 
strongly support the identification of \1wga \/ as a high
redshift cluster of galaxies, most probably at $z=0.85 \pm 0.15$, 
 with an inferred
temperature $kT =10^{+4}_{-3}$~keV  and an unabsorbed luminosity (in a $r=120^{\prime \prime}$ aperture) of 
$1.3^{+0.16}_{-0.14} \times 10^{45}$~erg s$^{-1}$ (0.5-10~keV).
This indication of redshift is also supported by the K and R band imaging, and 
is in agreement with the spectroscopic redshift of 0.89 found
by Ebeling et al. (2001).
The surface brightness profile is consistent with a
 $\beta$-model with $\beta = 0.770 \pm 0.025$, $r_c=(18.1 \pm
0.9)^{\prime \prime}$ (corresponding to $101 \pm 5$ kpc at $z=0.89$), and $S(0)=1.02 \pm 0.08$  counts arcsec$^{-2}$ .
\1wga \/ was selected as an extreme X-ray loud source with \fxv \/ $> 60$;
this selection method, thanks to the large area sampled,
  seems to be a highly efficient method
for finding luminous high z clusters of galaxies.

\end{abstract}

\keywords{
galaxies: high-redshift ---
galaxies: clusters: general ---
galaxies: clusters: individual (\1wga ) ---
X-rays: galaxies: clusters ---
X-rays: individual (\1wga )
}

\section{Introduction}

Clusters of galaxies are good tracers of the large scale
structure of the matter distribution in the universe.
The standard models of structure formation predict that
the cluster distribution and evolution is fully determined by the spectrum
of primordial perturbations and cosmological parameters $\Omega_0$ and 
$\Lambda$ (e.g. Press \& Schechter 1974; and later works), thus 
observations of high redshift clusters constrain these parameters 
(e.g., Oukbir \& Blanchard 1992). Moreover X-ray measurements of high 
redshift clusters of galaxies can place strong constraints on the
thermodynamic evolution of the intracluster medium (ICM). 
For example the Luminosity-Temperature ($L_X-T$) relation at 
different redshifts probes the interrelated evolution of
the cluster baryon mass and the total mass (e.g. Kaiser 1991; Evrard
\& Henry 1991; and later works).
Furthermore X-ray and observations of the Sunyaev-Zeldovich effect 
of a sample of objects at
different $z$ may be used to obtain an independent estimate of
$H_0$ (for a review see e.g., Birkinshaw 1999)

Given this potential wealth of information in the
past few years a great effort has been made  to search 
for high redshift clusters of galaxies (e.g. Rosati et al. 1998;
Vikhlinin et al. 1998).
Among the different methods, X-ray surveys  
have played the most important role, 
 but no method allowed the identification of more than
a handful of clusters at $z>0.8$.\\
The finding of clusters from the X-ray data was complicated mainly by the 
low spatial and/or spectral resolution of the previous X-ray missions.
{\it Einstein} and {\it ROSAT} marked an important step in the study of 
clusters but thanks to sub-arcsecond {\it Chandra} spatial resolution 
it is now possible to distinguish easily between point sources and the 
more diffuse X-ray emission from clusters at any redshift.

We present the {\it Chandra} observation of \1wga .
This source is one of 16 peculiar {\em ROSAT} PSPC sources
selected for their extremely high X-ray to optical flux ratio
(Cagnoni et al. 2000; Cagnoni et al., in prep.).
\1wga \/ is a  bright ($F_{0.1-2.4{\rm keV}} > 10^{-13}$ \cgs )
 WGACAT (White, Giommi \& Angelini
1994) source  with blank fields, i.e. no optical
counterparts on the Palomar Observatory Sky Survey to O=21.5.
The extreme \fxv \/ ratio that follows from this is incompatible
with all major and common classes of extragalactic sources, 
including normal quasars, AGNs, 
 normal galaxies and nearby clusters of galaxies (Maccacaro et
al. 1988).  Possibilities for the nature of these `blanks' 
(Cagnoni et al. in prep.)  include:

\begin{figure*}[tb]
\pspicture(0,2)(9,12)

\rput[tl]{0}(0.,12){\epsfxsize=8.2cm
\epsffile{fig1a.epsi}}

\rput[tl]{0}(10.5,12){\epsfxsize=7.2cm
\epsffile{fig1b.epsi}}

\rput[tl]{0}(-0.1,4.5){
\begin{minipage}{18.5cm}
\small\parindent=3.5mm
{\sc Fig.}~1.--- {\em Chandra} view of 1WGA~J1226.9$+$3332: (a) cleaned data
 (see text) in the whole S3 chip  and (b) zoom of the cluster
 (smoothed and background subtracted). 
\par
\end{minipage}
}
\endpspicture
\vspace{-0.5in}
\end{figure*}

%
(a) Quasar-2s, i.e.  high luminosity, high redshift
heavily obscured quasars, the bright analogs of 
Seyfert~2s; 
(b) Low Mass Seyfert-2s, that is AGNs powered by a low mass obscured
black hole (i.e. obscured Narrow Line Sy~1);
(c) AGNs with no big blue bump, e.g. ADAFs;
(d) Isolated Neutron Stars undergoing Bondi accretion from the
ISM (Madau \& Blaes 1994);
(e) $\gamma$-ray burst X-ray afterglows, 
(f) failed clusters, in which a large overdensity of matter 
has collapsed but has not formed galaxies  (Tucker, Tananbaum \& Remillard 1995); 
and, most relevant to this paper, (g) high redshift clusters of galaxies.  

Here we present strong evidence that \1wga \/ is indeed  a high redshift cluster.

We will use H$_0$=75~km~s$^{-1}$~Mpc$^{-1}$ and $q_0 = 0.5$;
errors in the paper  represent 1$\sigma$ confidence levels, unless 
explicitly stated otherwise.

\section{{\em Chandra} observations}

\1wga \/ was one of two `blanks' for which we obtained Cycle~1
{\em Chandra} (Weisskopf et al., 1996) observing time.  
It was observed in 
ACIS-S configuration (Garmire et al., in prep.) with the backside
illuminated S3 chip for a total useful exposure time of 9832.3 s.
The data was cleaned 
 as described in  Markevitch et al. (2000a) and for
the spectral and image analysis we used the latest available
ACIS background dataset (Markevitch et al. 2000b).

\begin{figure*}[b]
\pspicture(0,7)(4,12)

\rput[tl]{0}(3,12.1){\epsfysize=6.0cm
\epsffile{fig2.epsi}}

\rput[tl]{0}(0,5.5){
\begin{minipage}{18.5cm}
\small\parindent=3.5mm
{\sc Fig.}~2.--- Background subtracted surface brightness radial profile of
 \1wga \/ (error bars) together with the best fit  $\beta$-model (see text). 
\par
\end{minipage}
}
\endpspicture
\end{figure*}

\subsection{Spatial analysis}

To maximize the signal-to-noise ratio, 
we extracted the  image in the 0.5-5 keV band (Fig.~1a and b).
Along with a number of faint point-like   sources, 
\1wga \/ is clearly seen as an extended source on an arcminute scale.
It shows azimuthal symmetry, except for a possible excess to the
north-west, $\sim 40^{\prime \prime}$ from the cluster center. 

After subtracting the background, from the same regions in a normalized
ACIS background map  (Markevitch et al. 2000b), we extracted the X-ray surface
 brightness profile (Fig.~2) in concentric annular regions centered on 
the X-ray emission peak and chosen in order to have 20 counts per annulus.
The profile appears to be 
 smooth without any obvious central excess related to a cooling flow
 (Figs.~1 and 2).

We fitted the surface brightness profile with a standard $\beta$ model 
\footnote{$S(r)=S(0)[1+(r/r_c)^2]^{-3 \beta + 1/2}$}
(Cavaliere \& Fusco-Femiano  1976) using the {\em Sherpa} (Siemiginowska
et al., in prep.) modeling and fitting tool from the CXC analysis
 package CIAO 2.0 (Elvis et al., in prep.).
We obtained best fit values of $\beta = 0.770 \pm 0.025$, $r_c=(18.1 \pm
0.9)^{\prime \prime}$ (corresponding to $101 \pm 5$ kpc at $z=0.89$), and $S(0)=1.02 \pm 0.08$  counts arcsec$^{-2}$  with a  $\chi ^2$ of 27.5 for 26 degrees of freedom (d.o.f.).
The excess to the north-west of the cluster is also visible in the radial profile; a drop of the surface brightness at $\sim 40^{\prime \prime}$ is present in the radial profile for the north-west sector. Similar features in the surface brightness radial 
profiles were detected by {\em Chandra} in nearby clusters (e.g. Markevitch et al. 1999;2000a;2001, Vikhlinin et al. 2001; Mazzotta et al. 2001) and interpreted as signs of subclump motion.

\subsection{Spectral analysis}

We extracted an overall spectrum in a circle with $r=120^{\prime \prime}$ in  the 0.5-10 keV band in PI channels, corrected for the gain difference 
between the different regions of the CCD.  
The spectrum (Fig.~3) contains  $\sim 1100$ net counts and we binned it in
order to have  100 counts per bin\footnote{A smaller binning, e.g. as in Fig.~3, leads to similar results}. Both the effective area file (ARF) and
the redistribution matrix (RMF) where computed by weighting each position
dependent ARF's and RMF's by the X-ray brightness.
We fitted the spectrum  in the 0.5-10~keV range with an absorbed
Raymond-Smith model (Raymond \& Smith, 1977) using {\em Sherpa}.
The source redshift is treated as unknown and, because of the
the low statistics, no iron line or other line complex features
are expected to be visible. 

\begin{figure*}[tb]
\pspicture(0,2)(9,12)

\rput[tl]{0}(0.,12){\epsfxsize=8.5cm
\epsffile{fig3.epsi}}

\rput[tl]{0}(10.5,12){\epsfxsize=6.8cm
\epsffile{fig4.epsi}}

\rput[tl]{0}(-0.1,4.5){
\begin{minipage}{18.5cm}
\small\parindent=3.5mm
{\sc Fig.}~3.--- {\em Chandra} ACIS Spectrum and residuals to a Raymond-Smith 
model with $kT=10.23$ keV and $z=0.89$. The spectrum was binned, for display 
purposes, to obtain a minimum of 30 counts per bin.\\
\small\parindent=3.5mm
{\sc Fig.}~4.--- The contours of the fit with a Raymond-Smith
plasma at 1, 2 and 3 $\sigma$. 
The three blue solid lines represent the luminosity-temperature relation (Markevitch 1998) for $q_0 =0.5$ and the three blue dashed lines for $q_0 =0$.
The  red lines represent the limits on the redshift obtain from the R-K color
of the brightest object assuming it is a first ranked elliptical (Coleman, Wu \& Weedman 1980), while the  dotted lines are the limits from the magnitude of the first ranked elliptical (see text).  The big dot corresponds to the best fit to the {\em Chand
ra} spectrum.
\par
\end{minipage}
}
\endpspicture
\end{figure*}

In order to get an estimate of the temperature, we fixed
the equivalent hydrogen column density to the Galactic value
($N_H=1.38 \times 10^{20}$ cm$^{-2}$, Stark et al. 1992), 
the metal abundance to 0.3 times
 the Solar value, and we draw confidence  levels 
for the T and z (Fig. 4).
We find a  best fit value of
 $kT=10.24$~keV and  $z=0.85$ ($\chi ^2$ is 18.68 for 20 d.o.f.)\footnote{
Normalizing the background map using an area of \1wga \/  observation without sources (7.56\% lower background) consistent results are obtained:
kT=11.94 keV and z=0.87}.
However, as shown in Fig.~4, these values are not well constrained.

While submitting this paper we found that the same object had been
independently identified as a cluster of galaxies in the WARPS survey (Ebeling et al. 2001) and
observed for Sunyaev-Zel'dovich effect measurement by Joy et al. (2001).
The cluster spectroscopic redshift measured by Ebeling et al. (2001) 
is  $z=0.89$, which is within the errors of our estimate based on {\em Chandra} and optical/IR constraints (see Section~4). 
Fixing the redshift to this value we obtain a temperature of 
$kT =10.47^{+4}_{-3}$~keV (kT=12.07 keV using a normalized background), which is in good agreement  with that obtained from the Sunyaev-Zel'dovich measurement by Joy et al. (2001) ($kT =10.0^{+2.0}_{-1.5}$~keV).
The absorbed 0.5-2.0 keV flux\footnote{The main component in the flux error is the uncertainty on the cluster temperature (quoted in the text); using the $\pm 1\sigma$ values on T we derive an error of $+11$\% $-12$\% on the flux.
Other contributions come from the normalization chosen for the background ($\pm 9$\%), 
from the fraction of counts lost using the $r=120^{\prime \prime}$ aperture assuming a
 $\beta$-profile ($\pm 5$\%) and from the Poisson error on the counts ($\pm 3$\%)}
in a $r=120^{\prime \prime}$ aperture is $(3.0 \pm 0.3) \times 10^{-13}$~erg~cm$^{-2}$~s$^{-1}$, 
consistent with the Ebeling et al. (2001) PSPC measurement;
the 0.5-10.0 keV absorbed flux in the same aperture (Table~1) is $(8.03^{+0.96}_{-0.88}) \times 10^{-13}$~erg~cm$^{-2}$~s$^{-1}$. For z=0.89 the
 unabsorbed bolometric and 0.5-2.0 keV band luminosities are 
 $L=(2.2 \pm 0.2) \times 10^{45}$ erg~s$^{-1}$ and  $L_{X(0.5-2.0)}=(4.4 \pm 0.5) \times 10^{44}$ erg~s$^{-1}$ respectively.


\section{Optical, Infrared and Radio Observations}

We obtained an R-band image of the  \1wga \/ field on Feb. 2, 1997
using the SAO 1.2~m telescope on Mt.Hopkins. 
A R$ = 20.4 \pm 0.2$ galaxy is detected less than 1$^{\prime \prime}$
from the X-ray centroid (Table~1).

A K-band image of the field was obtained
using NSFCam at the  NASA IRTF on Jan. 31, 2000. 
The same galaxy is also seen, but is much brighter at
K=15.5 (Fig. 5) isophotal magnitude.
 The K-band  magnitude within the core radius of the X-ray source
is K=15.4, implying that little large scale IR emission
is present.
The R-K color of this object is  5.1 with an estimated uncertainty
of 0.3 mag due to the poor image quality in both K- and R-band.

Comparing R and K-band images, it is clear that there are many more objects
detected in the K- than in the R-band. We detect 
23 objects with K$<$19.5 mag in the $1.5'\times1.5'$ 
image using the source extractor software (SExtractor, 
Bertin \& Arnouts 1996) versus only 4 at R. 
The star/galaxy classification was carried out
morphologically using the Kron radius (Bertin \& Arnouts 1996), and 3  objects are 
classified as pointlike. These objects are also detected in the R-band  
and have  bluer colors (R-K$<$4.2) than the extended objects; we argue
they are stars.
We should not have to worry about the star contamination of
the sample since there are very few stars with K$>$16 (e.g. Glazebrook et al. 1995). 
The rest of the morphologically extended objects
are galaxies. These, including the bright one located
at the X-ray centroid, have very similar red colors ($4.5<$R-K$<6.5$), 
implying that they likely have 
 similar redshifts and belong to a cluster. 
A high-z cluster, 
CIG J0848+4453, at z=1.27 was detected in a near-IR field survey using 
 similar color criteria in an over-dense region (Stanford et al 1997).
The apparent asymmetry of the galaxy distribution may be an artifact 
of the longer exposure time at the center of the K-band mosaic image which is offset from the X-ray centroid.

The galaxy surface density in the K-band image is clearly high (Fig.~5). 
We generate K-band galaxy number counts in our field in the  range
$15<K<19.5$, and compare them with those obtained from
the near-IR field survey in this  range (Gardner et al. 1993,
Saracco et al. 1997, Huang et al. 1997, Minezaki et al. 1998, 
Huang et al. 2000)  with their coverage ranging from 200~arcmin$^2$ to
10~deg$^2$. The number counts obtained 
from our image are substantially higher ($> 10 \sigma$, a factor of 
100 at K=19) than those from the field
surveys. Such a large excess cannot be due to  Poisson statistics or magnitude
errors.

Both the high density and  the similar (extremely red) colors for these galaxies thus  imply that they are likely to be members of a cluster.

The FIRST survey detected a  faint $3.61 \pm 0.18$~mJy source
 at 1.4 GHz  (Becker, White \& Helfand, 1995) close to 
the center of the X-ray emission (Table~1).
The radio source is pointlike (FWHM$ \leq 0.91^{\prime \prime}$, $\leq 16$~kpc at $z=0.85$) 
and has a luminosity $L_R (z=0.85) \sim 6.6 \times 10^{24}$ ~W~Hz$^{-1}$, 
compatible with low luminosity radio-loud AGNs (e.g. Zirbel \& Baum, 1995).

\section{Discussion and Conclusion}

{\em Chandra} has shown that \1wga \/ is an extended X-ray source,
with a hard, high temperature thermal spectrum.
Optical and IR imaging has shown that  \1wga \/  has 
faint optical/IR counterparts. 
A cluster of galaxies is
the only known type of object that could fit such a
description. Moreover the K-band image   shows a strong excess
of galaxies compared with field counts at K$ > 17$ (Fig.~ 5) around \1wga .
 Physical clustering is the only possible explanation.
Several lines of argument go on to suggest that it
 is a high redshift cluster of galaxies.
Below we list these arguments and  try to constrain its
 temperature and redshift. These results are summarized in Figure~4.

\noindent
(1)
The {\it Chandra} X-ray profile is well fitted by a $\beta$ model with
$\beta$=0.77, a value in agreement with a typical relaxed cluster (e.g.
Jones \& Forman, 1999 and references therein).
Moreover, if the cluster redshift is $0.7<z<1.2$, then  
the observed angular core radius, $r_c=(18.1 \pm
0.9)^{\prime \prime}$ ($101 \pm 5$ kpc at $z=0.89$ with the assumed 
cosmology),  
corresponds to a linear size of $90<r<150$~kpc 
for any value of $\Omega$ ($130<r<220$~kpc 
for any value of $\Omega$ for $H_0$=50~km~s$^{-1}$~Mpc$^{-1}$),
values  consistent with  a typical relaxed cluster.

\noindent
(2) 
It is well known that clusters of galaxies follow 
a well-defined Luminosity-Temperature 
relation (e.g. Markevitch 1998 and reference therein).
Recently it has been shown that the local $L_X-T$ relation does not evolve
(or is consistent with little evolution) with redshift up to $z\approx 0.8$
(see e.g. Wu et al. 1999; Della Ceca et al
2000; Fairley et al. 2000 and reference therein).
Fig.~4 shows that the $L_X-T$ relation ($L = A T_6^{\alpha}$ where 
$T_6 =T/6$ keV,  $\alpha =2.02 \pm 0.40$ and 
$A=(1.71 \pm 0.21) \times 10^{44}$ $h^{-2}$ erg s$^{-1}$)
 from Markevitch (1998) requires a 
$3 \sigma$ lower limit of $T>4$~keV and  z$>$0.4, while
the $1 \sigma $ limits require T$>7.5$~keV and $z>0.65$ 
if no evolution is assumed.
The best fit value of 0.85 obtained from {\em Chandra} spectrum (dot in
Figure~4) is  consistent with the unevolving $L_X-T$ relation.

\begin{figure*}[tb]
\pspicture(0,2)(9,12)

\rput[tl]{0}(4.5,12){\epsfxsize=8cm
\epsffile{fig5.epsi}}

\rput[tl]{0}(2,4.5){
\begin{minipage}{18.5cm}
\small\parindent=3.5mm
{\sc Fig.}~5.--- K-band image of \1wga . The circle shows the {\em Chandra}
 X-ray core radius of $18^{\prime \prime}$.
\par
\end{minipage}
}
\endpspicture
\vspace{-0.8in}
\end{figure*}

\noindent
(3) 
K-band imaging has shown that \1wga \/ has
a galaxy at K=15.5. If this source is a first ranked cluster elliptical
with $M_K =-26.7 \pm 0.5$ (Collins \& Mann 1998),  
 then, assuming negligible K-correction,  $q_0 =0.5$ and , H$_0$=50~km~s$^{-1}$~Mpc$^{-1}$ as in  Collins \& Mann (1998),
it has  $z=0.68^{+0.30}_{-0.19}$ (dotted lines in Fig.~4).

\noindent
(4)
the  color  of the K-band galaxies is extremely  red, with none bluer than R-K$\sim 4.5$ and a maximum R-K$=6.5$. 
Only a unevolving elliptical galaxy at $0.7<$z$<1.5$  
or a Sbc galaxy at z$>1.1$ can have such a red color (Coleman, Woo \& Weedman, 1980).
Using the more accurate R-K$\sim 5.1 \pm 0.3$ of the first ranked elliptical 
we can restrict the redshift range to $0.75 < z < 1.0$ 
(red lines in Fig.~4).

Using all these constraints we conclude that 
\1wga \/ is a distant  cluster of galaxies with a most probable  redshift of 
$0.85 \pm 0.15$ and not  smaller than $z=0.65$.

This  gives $kT =10^{+4}_{-3}$~keV and implies an X-ray luminosity, determined in a $r=120^{\prime \prime}$ aperture, of 
$L_{X (0.5 - 10 keV)} = 1.3^{+0.16}_{-0.14} \times 10^{45}$ erg s$^{-1}$, 
 corresponding, for $z=0.85$,  to a bolometric  $L=(2 \pm 0.2) \times 10^{45}$ erg s$^{-1}$.
Our estimated redshift is similar to the spectroscopic $z=0.89$ found by Ebeling et al. (2001).

The blank field X-ray source  \1wga  \/ is thus  a highly luminous and 
massive high redshift cluster and a useful source to
determine the evolution of the clusters X-ray luminosity function (e.g. Rosati et al., 1998).
Since models in the direction of low $\Omega$ universe 
(with or without cosmological constant) (e.g. Henry 2000, Borgani \& Guzzo 2001) predict a higher density of high redshift clusters compared to high $\Omega$ models, finding hot high redshift clusters has a strong leverage on cosmological models.

Since such high luminosity, high redshift clusters should be rare, 
the relative ease with which 
this discovery was made is potentially of great significance.  
The search for high \fxv \/ sources (`blanks'), sampling a large area of the sky, is an efficient method of finding very luminous high redshift clusters;
serendipitous typical flux limited surveys can find (and found) plenty of $z>0.6$
clusters, but they are inefficient at finding such luminous clusters
because they cover relatively small areas ($\sim 100$ deg$^2$). 
This methodology is a useful complement to serendipitous flux limited surveys.

\acknowledgments

We gratefully acknowledge the work by the whole Chandra team in
making {\em Chandra} a great success and the staff of CXC for the rapid reprocess 
of the data and the CIAO data analysis software.
We are grateful to M. Markevitch for making his data available in electronic form  and for his prompt answers to ACIS background related questions.
We thank M. Massarotti, R. Della Ceca, M. Chiaberge, S. Borgani and S. Andreon 
for useful discussion.
IC thanks A. Fruscione, F. Nicastro and  A. Siemiginowska for a quick
introduction to CIAO. 
We are also grateful to the staffs of NASA-IRTF and FLWO. 
 
This research has made use of the
NASA/IPAC Extragalactic Database (NED) which is operated by the Jet
Propulsion Laboratory, CalTech, under contract with
the NASA. 

This work was supported by NASA grant GO 0-1086X
 and by the Italian MURST (IC and AC).
P.M. acknowledges an ESA fellowship.

{\footnotesize
\renewcommand{\arraystretch}{1.4}
\renewcommand{\tabcolsep}{1mm}
\begin{center}
TABLE 1
\vspace{1mm}

{\sc  1WGA~J1226.9$+$3332 observations}
\vspace{1mm}

\begin{tabular}{cccccccc}
\hline \hline
Energy Band & Instrument	& Date	&Exposure	&Coordinates	&Offset$^a$	&Count Rate	&Flux (units)\\
	    &			&	&(ks)		&(J2000)	&($\prime \prime$) &(Counts s$^{-1}$)	& \\
\hline
X-ray	&{\it Chandra}		&Jul 31 2000	&9832.3	&12 26 58.2 +33 32 48.28  &0.0 	&$0.107 \pm 0.006^b$ 	&$8.03 \times 10^{-13}$ (\cgs $^c$)\\
R-band	&SAO 48in		&Feb 02 1997	&900	&12 26 58.2 +33 32 48.7	  &0.87	&--			&$20.4 \pm 0.2$ (Mag.)\\
K-band	&IRTF			&Jan 31 2000	&1200	&12 26 58 +33 32 48$^c$   &2.0	&--	&15.5 (Mag.)\\
Radio	&FIRST			&--		&--	&12 26 58.19 +33 32 48.61 &0.79	&--			&$3.61 \pm 0.18$ (mJy at 1.4~GHz)\\
\hline
\end{tabular}
\end{center}
$^a$ Offset from {\em Chandra} position\\
$^b$ [0.5-10 keV] in a circle with r=120$^{\prime \prime}$ computed from the spectral model within sherpa\
$^c$ Due to the lack of bright stars in the K-band image, 
we obtained an estimate of the position using the objects in common with the R
image.\\
}
\vspace{3mm}

\end{document}